\documentclass[aps,prd,epsf,amsmath,amssymb,graphics,11pt]{revtex4} %
\usepackage{epsfig}
\usepackage{graphicx}
\usepackage{dcolumn}
\usepackage{bm}
\usepackage{xcolor}

\newcommand{\eq}[1]{\begin{equation} #1 \end{equation}}

\newcommand{\vev}[1]{\left<#1\right>}

\newcommand{\ovaln}[1]{\left(#1\right)}
\newcommand{\pravok}[1]{\left[#1\right]}

\begin{document}
  \title{Dark Matter from Q4 Extension of Standard Model}
\author{I. Lovrekovi\'c\footnote{e-mail: {\tt iva.lovrekovic@gmail.com}}}
\affiliation{Institut f\"ur Theoretische Physik, Technische Universit\"at Wien,
 Wiedner Hauptstr.~8-10/136, A-1040 Wien, Austria}
   \begin{abstract}      

 Recent analysis of spontaneous breaking of SU(2) as a continuous gauged flavor symmetry to non-abelian discrete group $Q_{4}$ as a residual symmetry motivates its consideration as a group for stabilizing dark matter.  We  determine the region of hidden sector particles compatible with relic dark matter density and present prospects in which these particles could be observed using direct and indirect dark matter search experiments.
    \end{abstract}

\maketitle     

\section{Introduction}
Closeness of Gauge-couplings at the Unification scale, neutrino masses, structure of lepton mixing matrix and hierarchy of quark mixing angles,  dark matter, dark energy and baryon asymmetry of the Universe are some of the still unexplained experimental facts. 
While framework of solving them all  doesn't exist yet, we can pick one issue and offer possible solution which may lead towards solving more of them. 
This can be done on different levels: choosing new fields and adding them within appropriate representation,   extending the gauge symmetry i.e. following the gauge route, adding supersymmetry or extending and modifying basic concepts like quantum fields and nature of space-time, increasing levels of speculation with each step we take. Here we will focus on the search for the dark matter (DM). 

 There are many evidences of DM existence \cite{Bergstrom:2012fi} and reasons for expecting it to be of particle nature.  Such a particle would have to be neutral, stable over cosmological scales and provide current DM relic density. It is reasonable to assume that the group used for accommodation of DM and SM extension   simultaneously describes more observations. Special structures of the lepton mixing matrix  and the quark mixing matrix  give quite a strong hint for a flavor symmetry $G$  broken in a non-trivial way. Requirement that at least two fermion generations
can be unified by $G$ 
leads us to the conclusion that the best choice for $G$  is a discrete, non-abelian symmetry.  \cite{Blum:2007jz}  
\newline
Since the discrete symmetry that stabilizes dark matter must be a remnant of a spontaneously broken gauge symmetry \cite{Krauss:1988zc}, in most cases the used symmetry is $U(1)_{X}$ gauge symmetry which is broken to discrete $Z_{2}$ subgroup.
Preservation of $Z_{2}$ symmetry stabilizes hidden sector in which the lightest particle is usually DM candidate. Examples of $Z_{2}$ symmetry are R-parity in Minimal Supersymmetric Standard Model  and T-parity in Little Higgs Model  \cite{Belyaev:2006zz}.

Simplest way to analyze extension of SM with particles in hidden sector stabilized with $Z_{2}$ symmetry is to  add one singlet scalar to SM particles which has been discussed in \cite{Burgess:2000yq}. 

 However, DM can as well consist of more particle types. In this case $U(1)_{X}$ gauge symmetry is  broken to  group  $Z_{N}$ with $ N>2$  and fields associated to its  irreducible representations create hidden sector of the model. Stabilization of DM with $Z_{3}$ symmetry has been studied in  \cite{Ma:2007gq}  where the DM candidate was neutral complex scalar singlet,  while DM phenomenology with discrete $Z_{2}$, $Z_{3}$  and $Z_{4}$ stabilizations of one extra sclalar doublet and one extra complex scalar singlet was investigated in \cite{Belanger:2012vp}.
In \cite{Adulpravitchai:2011ei}, a non-Abelian discrete group $D_{3}$ was recently employed to stabilize hidden sector consisting of one or two component DM.

Here we investigate the stabilization of one component DM with quaternion group, $Q_{4}$. It has been shown recently that $Q_{4}$ group is the only non-Abelian discrete group which arises from spontaneous breaking of continuous (possibly gauged) flavor  symmetry SU(2) \cite{Adulpravitchai:2009kd}.
It is a double valued counterpart of the dihedral group $D_{2}$, also called $D_{2}'$,  and it has been used in series of articles by M. Frigerio and collaborators \cite{Frigerio:2004jg}, \cite{Frigerio:2005pz}, \cite{Frigerio:2007nn} for new understanding of family symmetries, $V_{CKM}$ and $U_{MNSP}$ matrices and correlation of $\Delta m_{12}^{2}$\text{ to }$\theta_{23}$ .

We assign new particles to non-trivial  irreducible representations of the group and stabilize the hidden sector  preserving the symmetry after electroweak symmetry breaking. We add two complex scalars in a doublet and three real scalars in singlet representations and we can compare the model with one component case of DM stabilization with $D_{3}$, $Z_{3}$ and $Z_{2}$.
With that choice of particles the most similar model turns out to be the one based on $D_{3}$ symmetry and in this article we follow closely their analysis of one-component DM model.
 In section II we present the group $Q_{4}$, its irreducible representations and decomposition of tensor products. Section III introduces potential for DM in $Q_{4}$ model, conditions on the model to achieve potential stabilization and new processes which take place in the early Universe.  It also shows the potential terms and reactions which differentiate the $Q_{4}$  from the $D_{3}$ DM stabilization. Considering  annihilation reactions, decays and DM conversion we use micrOMEGAs package \cite{Belanger:2006is},\cite{Belanger:2008sj},\cite{Belanger:2010gh} to  present its mass dependence on  the strength of Higgs portal coupling and investigate prospects of direct and indirect detection of lightest DM particle in this model. In section IV we conclude.

\section{$Q_{4}$ group}

$Q_{4}$  contains eight elements $a^{m}b^{k}$ for  $m$=0, 1, 2, 3 and $k=$0, 1, four one dimensional irreducible representations  $\text{\bf{ 1}}_{++}$, $\text{\bf{ 1}}_{+-}$,
$\text{\bf{ 1}}_{-+}$, $\text{\bf{ 1}}_{--}$ ($\equiv\text{\bf{1}}_{i}, i=1,...,4$ respectively) and one pseudo-real two dimensional representation \text{\bf{2}}. The generators of one dimensional representations are $a=1$, $b=1\text{ for }\text{\bf{1}}_{1},  
a=1, b=-1$  for
 $\text{\bf{1}}_{2}$; $a=-1$, $b=-i$  for $\text{\bf{1}}_{3}$; $a=-1$, $b=i$  for $\text{\bf{1}}_{4}$ and for two dimensional representation they are
$a=\ovaln{\begin{array}{cc}e^{\frac{\pi i}{2}} & 0 \\ 0 & e^{-\frac{\pi i}{2}}\end{array}}$\hspace{-0.05cm}\text{, }
$b=\ovaln{\begin{array}{cc}0 & i \\ i & 0 \end{array}},$\hspace{-0.1cm}
\text{ satisfying } $a^{2}=b^{2}$, $aba=b$. 
Since the two dimensional representation is pseudo-real there exists a similarity matrix $U$ between the transformation matrix and its complex conjugate, such that $$\ovaln{\begin{array}{c}-a_{2}^{*}\\a_{1}^{*}\end{array}}=U\ovaln{\begin{array}{c}a_{1}^{*}\\ a_{2}^{*}\end{array}}$$ \text{doublet transforms as} $\ovaln{\begin{array}{c}a_{1}\\a_{2}\end{array}}\sim\bf{2}$, \text{ where } $U=\ovaln{\begin{array}{cc}0 & -1 \\ 1 & 0\end{array}}$. 

The tensor product of two doublets 
$\ovaln{\begin{array}{c} a_{1} \\ a_{2} \end{array}}\times \ovaln{\begin{array}{c} b_{1} \\ b_{2} \end{array}}$
decomposes in \eq{\begin{array}{cc}(a_{1}b_{2}-a_{2}b_{1})_{1_{1}}&
(a_{1}b_{2}+a_{2}b_{1})_{1_{4}}\\ 
(a_{1}b_{1}-a_{2}b_{2})_{1_{2}} & (a_{1}b_{1}+a_{2}b_{2})_{1_{3}}\end{array},\label{decomp22}}
products  of singlets with doublet in
\eq{\begin{array}{c}
(\omega)_{1_{1}}\times\ovaln{\begin{array}{c} a_{1} \\ a_{2}\end{array} }_{2}=\ovaln{\begin{array}{c}\omega a_{1} \\ \omega a_{2}\end{array}}_{2},\\
(\omega)_{1_{4}}\times\ovaln{\begin{array}{c} a_{1} \\ a_{2}\end{array} }_{2}=\ovaln{\begin{array}{c}\omega a_{1} \\ -\omega a_{2}\end{array}}_{2},\\ 
(\omega)_{1_{2}}\times\ovaln{\begin{array}{c} a_{1} \\ a_{2}\end{array} }_{2}=\ovaln{\begin{array}{c}\omega a_{2} \\ \omega a_{1}\end{array}}_{2},\\
(\omega)_{1_{3}}\times\ovaln{\begin{array}{c} a_{1} \\ a_{2}\end{array} }_{2}=\ovaln{\begin{array}{c}\omega a_{2} \\ -\omega a_{1}\end{array}}_{2}.\end{array}}
The tensor products of singlets with singlets in
\eq{1_{s_{1}s_{2}}\times1_{s'_{1}s'_{2}}=1_{s''_{1}s''_{2}}\label{dec11},} 
where $s_{1,2}\in\{+,-\}$\text{, } $s''_{1}=s_{1}s'_{1}$ and $s''_{2}=s_{2}s'_{2}$.  Dihedral groups and their double covers are nicely described in \cite{Blum:2007jz}, \cite{Ishimori:2010au}.

\section{Stabilization of dark matter}

We want to find the best way to arrange particles in $Q_{4}$ irreducible representations to differentiate from DM stabilization with $Z_{2}$, $Z_{3}$ and $D_{3}$ groups.
It is clear from the multiplication rules that Lagrangian extended with particles from singlet representations preserves additional $Z_{2}$ symmetry. 
 To address a nontrivial case we add real particles $\eta_{i}$ i=2,3,4  in each nontrivial singlet representation and complex  $X=\ovaln{\begin{array}{c} \chi \\ \chi^{*} \end{array}}$  in a  doublet of $Q_{4}$  which amounts to five new degrees of freedom.
Scalar sector of the Lagrangian for DM  stabilized with Q4 symmetry is invariant under $Q_{4}$ and gauge symmetries and beside SM Higgs field, it has kinetic terms and potential. It can be constructed using multiplication rules of singlets and doublet eq.(\ref{decomp22})-(\ref{dec11}) and reads 
\eq{
\begin{array}{rl}V&=m_{1 }^{2}(H^{\dagger}H)+m_{2i}^{2}\eta_{i}^{2}+\frac{\mu}{2}(X^{\dagger}X)+\alpha_{12}\eta_{2}(X^{\dagger}X)_{2}+\alpha_{13}\pravok{\eta_{3}(X^{\dagger}X)_{3}+H.c.} \\ &+\alpha_{22}\pravok{\eta_{2}(XX)_{2}+H.c.}+\alpha_{23}\eta_{3}\ovaln{XX}_{3}+\alpha_{24}\eta_{4}\ovaln{XX}_{4} +\alpha_{3i}\eta^{2}_{i}(X^{\dagger}X)+\alpha_{4i}\eta_{i}^{2}(H^{\dagger}H)\\ & +\alpha_{5}(H^{\dagger}H)(X^{\dagger}X) +\lambda_{1}(H^{\dagger}H)^{2} 
+\lambda_{i+1}(X^{\dagger}X)_{i}(X^{\dagger}X)_{i}+\lambda_{5}\pravok{(XX^{\dagger})_{2}(XX)_{2}+H.c.}\\ &+\frac{\lambda_{6i}}{4}\eta_{i}^{4}\end{array}
\label{pot}}

\noindent Where $\alpha_{1i}$ coupling exists for i=2,3 while the term with $\alpha_{14}$ vanishes.

The symmetry is preserved after electroweak symmetry breaking 
$\vev{H}=\frac{1}{\sqrt{2}}\ovaln{\begin{array}{c}0 \\ v \end{array}}, \vev{\eta}=0,\vev{\chi}=0 $
where $v^{2}\equiv-m_{1}^{2}/\lambda_{1}\approx(246\text{ GeV})^{2}$ is squared Higgs vacuum expectation value. 
For simplicity we assume CP conservation, so all the couplings in potential are real.  
Phenomenological constraints on the couplings come from demanding that the potential is bounded from  below and that the electroweak vacuum is the global minimum of the potential. This gives the conditions on the terms with quartic couplings which must be positive, while the terms with $\alpha_{1i},\alpha_{2i},\alpha_{3i},\alpha_{4i}$ for $i=2,3,4$ and $\alpha_{5}$ must be greater than some minimum (negative) value. The parameter $\mu$ is determined by the condition $m_{\chi}^{2}>0$.

A model based on the $Q_{4}$ symmetry has all the terms in the Lagrangian like the model based on the $Z_{3}$ symmetry besides the term $\chi^{3}+\chi^{*3}$. The theory based on an Abelian discrete $Z_{3}$ symmetry is most similar to the recent one based on the group $D_{3}$. The minimal theory based on the $D_{3}$ symmetry has nontrivial interaction predicted by the term $\frac{i}{3!}\eta\ovaln{\chi^{3}-\chi^{*3}}$ which allows the decay $\eta\rightarrow3\chi3\chi$. Where $\eta$ transforms as a nontrivial singlet and $\chi$ as a doublet of irreducible representations of $D_{3}$ group. 
New terms in the theory based on the $Q_{4}$ symmetry 
are those with coupling $\alpha_{1i}$ and $\alpha_{2i}$ which allow the decay of $\eta_{i}$ particles. For $i=2,3$ the decays are into $\chi\chi$ and $\chi^{*}\chi^{*}$ and for $i=4$ into $\chi^{*}\chi$. There are also new terms  that govern reactions which don't contribute to change in number and species of DM particles.

After expansion $v\rightarrow v+h$ around the vacuum from potential we can read these physical masses:

\eq{\begin{array}{l}
m_{h}^{2}=2 \lambda_{1}v^{2}\\
m_{\eta_{i}}^{2}=m_{2i}^{2}+\alpha_{1i}v^{2}\\
m_{\chi}^{2}=\frac{\mu}{2}-\alpha_{5}v^{2}
\end{array}}

To examine the thermal relic abundance of added particles in $Q_{4}$ model we use the micrOMEGAs package \cite{Belanger:2006is},\cite{Belanger:2010gh},\cite{Belanger:2008sj}. Relic abundance of the particles is governed by Boltzmann equations which account for the change of the total number of particles we are interested in.

From five new added particles the lightest one is taken to be our DM candidate.  If the decays $\eta_{2,3}\rightarrow\chi\chi$, $\eta_{2,3}\rightarrow\chi^{*}\chi^{*}$, $\eta_{4}\rightarrow\chi^{*}\chi$ are kinematically allowed this is $\chi$ particle. If these decays were not allowed there would be five DM candidates  and calculation of relic density would have to be treated using five coupled Boltzmann equations. This number reduces to four because of CP invariance of the potential (\ref{pot}) $n_{\chi}=n_{\chi^{*}}$ so one needs to solve only Boltzmann equation for $n_{\chi}$ and total relic density becomes $n_{\chi+\chi^{*}}=2n_{\chi}$ giving the relic abundance $\Omega h^{2}=\Omega_{\chi+\chi^{*}}h^{2}$.  Under certain circumstances 
when the second-lightest particle is only slightly heavier than the lightest one  if certain  condition (see (6) below) is guaranteed two Boltzmann equations can  be reduced to  obtain one 
and it is possible to calculate relic density via standard methods \cite{D'Eramo:2010ep}. In this case  there can be presence of co-annihilations if the model allows it. 
Reactions which contribute to Boltzmann equations in calculation of relic density are
\begin{description}
\item{(a)}  \textit{annihilations into SM particles}: $\eta_{i}\eta_{i}\rightarrow X_{SM}$, $\eta_{i}\eta_{j}\rightarrow X_{SM}$, $\chi^{*}\chi\rightarrow X_{SM}$, 
 $\chi\chi\rightarrow X_{SM}$
\item{(b)}  \textit{semi-annihilations}: $\chi\eta_{2,3}\rightarrow\chi^{*}H$, $\chi\eta_{4}\rightarrow\chi H$, $\chi\chi\rightarrow H\eta_{2,3}$,  $\chi^{*}\chi\rightarrow H\eta_{4}$ 
\item{(c)}  \textit{DM conversion}:
\begin{description}
\item{(c1)} \textit{dark annihilations:}
 $\eta_{i}\eta_{i}\rightarrow\eta_{j}\eta_{j}$,  $\eta_{i}\eta_{i}\rightarrow\chi^{*}\chi$,                
$\chi^{*}\chi\rightarrow\eta_{i}\eta_{i}$, $\chi^{*}\chi\rightarrow\eta_{2}\eta_{3}$, \newline $\chi\chi\rightarrow\chi^{*}\chi^{*}$,  
$\chi\chi\rightarrow\eta_{2,3}\eta_{4}$,  
\item{(c2)} \textit{dark co-annihilations:}
$\eta_{4}\chi\rightarrow\eta_{2,3}\chi^{*}$,  $\eta_{2,3}\chi\rightarrow\eta_{4}\chi^{*}$, 
$\eta_{2,3}\eta_{4}\rightarrow\chi\chi$,  $\eta_{2,3}\eta_{4}\rightarrow\chi^{*}\chi^{*}$, \newline $\eta_{2}\eta_{3}\rightarrow\chi^{*}\chi$
\end{description} 
\item{(d)}  \textit{decay}: $\eta_{2,3}\rightarrow\chi\chi$, $\eta_{2,3}\rightarrow\chi^{*}\chi^{*}$, $\eta_{4}\rightarrow\chi^{*}\chi$
\end{description}

The reactions with $\chi$ are analogous to reactions with $\chi^{*}$ so we list only those for $\chi$ and $\eta_{i}$.
Quaternion symmetry allows  reactions of annihilation into SM particles (a), semi-annihilation (b) recently studied in \cite{D'Eramo:2010ep},  DM conversion (c) among which we recognize "dark" co-annihilations and "dark" annihilations, and decay (d).  In this work we focus on reactions of type (a) with reactions (d) necessary included to make sure we have one   DM candidate. DM conversion reactions in $D_{3}$ model were shown to   change the number of heavier particles while they didn't influence the number of lightest DM particle or change the relic density of lightest DM particles for chosen set of parameters. The lightest DM particle would be influenced by these type of reactions if the masses of the heavier and the lighter particle were nearly degenerate \cite{Adulpravitchai:2011ei}. 

\vspace{0.5cm}

\textit{Annihilations.}
Reactions of annihilation of DM particles into SM particles are common to scalar DM models with Higgs portal interaction.  In the case $m_{\chi,\eta_{i}}<<m_{H}$  annihilation cross-section depends on the decay width of SM Higgs boson in all kinematically allowed final states $X_{SM}=ZZ,WW,b\overline{b}...$ of SM and  of the hidden sector ('dark' annihilations).  The decaying Higgs boson is a virtual particle  with a mass of two times the mass of annihilating particles. 
The cross section of annihilating particles in the region $m_{H}>>m_{\chi},\eta_{i}$ reduces to \cite{Burgess:2000yq}  $\left< \sigma v\right>_{\chi\rightarrow X_{SM}}\approx\frac{\alpha_{5}^{2}}{4\pi m_{\chi}^{2}}$ depending primarily on $\alpha_{5}$ coupling and DM mass $m_{\chi}$.
To determine the value of Higgs portal coupling to $\chi$ we use the requirement that $\chi$ gives the right relic abundance within the WMAP observations $0.09<\Omega h^{2}<0.13$ like done in \cite{Boucenna:2011tj} while we  assume $\eta_{2,3,4}$  have already decayed earlier to lighter particles.  

\begin{figure}
 \moveleft0.08\linewidth\vbox{
\includegraphics[width=1.16\linewidth]{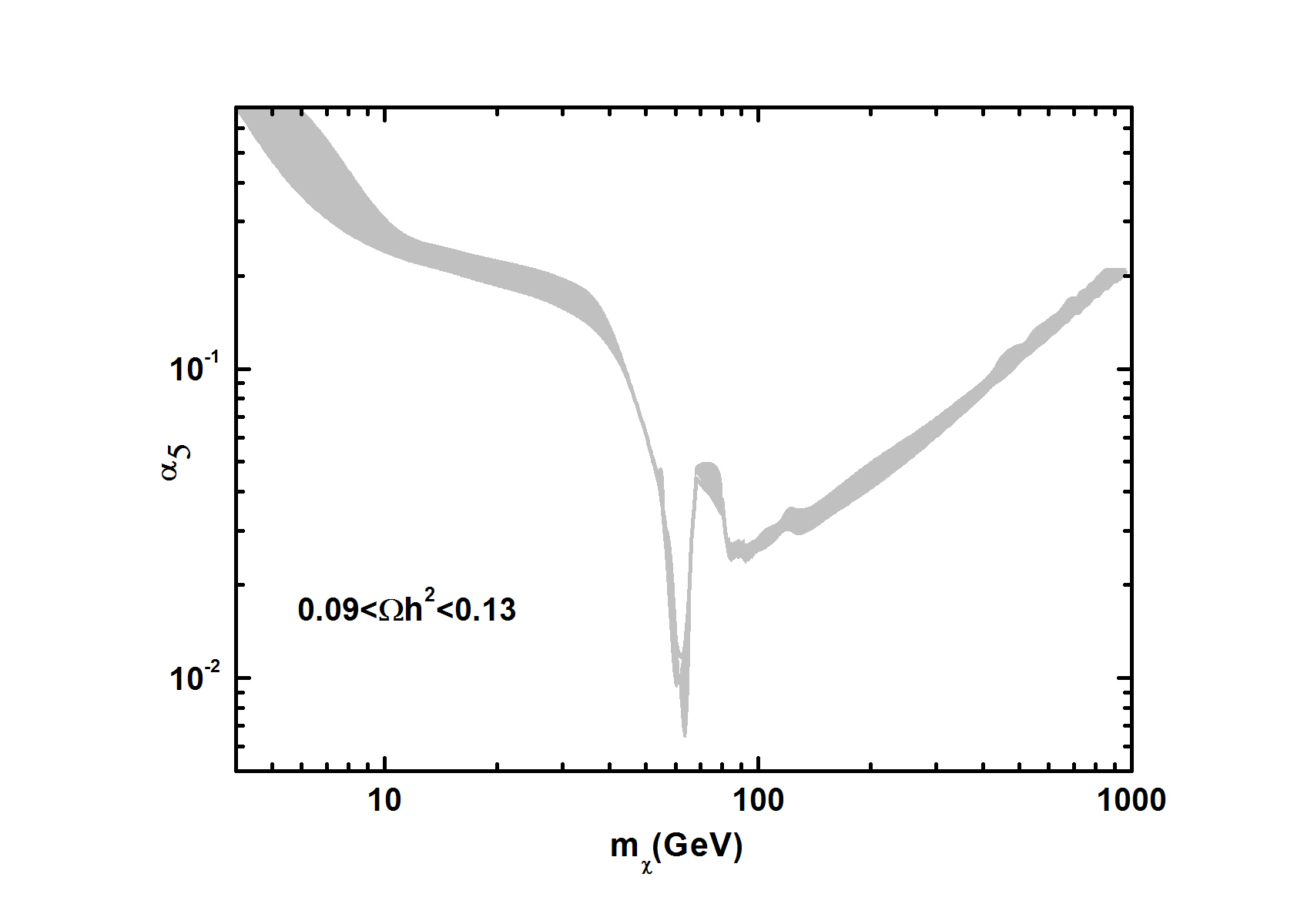}    
}
\caption{ Dependence of $\alpha_{5}$ coupling on the masses of DM particle $m_{\chi}$. Along the band values of relic density are within the WMAP allowed region $0.09<\Omega h^{2}<0.13$. The region below is excluded $\Omega>\Omega_{\text{WMAP}}^{\text{DM}}$. The resonance at $m_{\chi}\approx\frac{m_{H}}{2}$ allows for $\alpha_{5}$ value to be of order of $10^{-3}$.
}
	\label{alph5_mchi}
\end{figure}

The  dependence of $\alpha_{5}$ on $m_{\chi}$ for the range of the $m_{\chi}$ from 4 GeV to 1000 GeV, often measured in experiments searching for DM,  is shown on the Figure \ref{alph5_mchi} in the  $m_{\chi}-\alpha_{5}$ plane for 
mass of the $\eta_{2}$ particle $\geq$ 2TeV so that reactions (d) are allowed for every $m_{\chi}$. 
 The  band shows values for which relic abundance is within the WMAP values. In the region under the band
$\Omega>\Omega_{\text{WMAP}}$ and above the band there have to be other reactions  strong enough to source the relic abundance to give correct  $\Omega.$  The strength of the coupling $\alpha_{5}$ in the region around Higgs resonance $m_{\chi}\approx\frac{m_{h}}{2}$ is very weak compared to the cases where $m_{\chi}<<\frac{m_{H}}{2}$ and $m_{\chi}>>\frac{m_{H}}{2}$  and  $\chi$ particles annihilate there very efficiently. In this region the value of $\alpha_{5}$  does not have to be very strong to account for the viable cosmology. 

To analyze dependence of annihilation channel $\chi^{*}\chi$ on  masses $m_{\eta}$ and $m_{\chi}$  
 we observe dependence $\alpha_{5}-m_{\chi}$ for $m_{\eta_{i}}\text{, }m_{\chi}<<m_{H}$ and $m_{\eta_{i}}\text{, }m_{\chi}>>m_{H}$ with i=2, 3, 4. To be sure reactions (d) are  allowed we keep $m_{\chi}<2m_{\eta_{i}}$. This choice of parameters for both regions gives good agreement with the graph on Figure (\ref{alph5_mchi}).  Reason for this is that reactions of the type (a) strongly dominate over (c) .
For example, cross-section of reactions $\chi\chi^{*}\rightarrow\eta_{i}\eta_{i}$ is of the order  of $\sim10^{-13}$ pb for center of mass momentum 500 GeV while dominant SM channel for this set of parameters $W^{+}W^{-}$ is $\sim10^{-2}$ pb for $m_{\chi},m_{\eta_{i}}>m_{H}$ region (for $m_{\chi}$=170 GeV and $m_{\eta_{i}}\approx 400$ GeV).
\color{black}

 \vspace{0.5cm}
\textit{Dark matter conversion processes} are reactions in which number of dark matter particles remains the same. 
There are two types of processes.  The "dark" annihilation reactions (c1) are $2\rightarrow2$ processes in which the number of DM particles of a particular species changes by two units. The reactions (c2) can be regarded as dark co-annihilations.  In these reactions the DM particle of one species annihilate with DM particle of another species.

If we had two hidden sectors each belonging to its particle type,  allowed decay of the second lightest particle to the lightest one would make the lightest one  DM candidate forming one-component DM. If this decay doesn't exist or it is  not kinematically  allowed, we have two-component DM and two hidden sectors, with relic density of each of them obtained by calculating two coupled Boltzmann equations.
 However, if the mass of the second lightest particle becomes smaller than $1.5$ times the mass of the lightest one \cite{Belanger:2006is}, \cite{Gondolo:1998ah} the inclusion of co-annihilations  lowers the relic density and makes possible the simple use of one Boltzmann equation. 
This is due to possibility of introducing the condition
\eq{\frac{n_{i}}{n}\approx\frac{n_{i}^{eq}}{n^{eq}}}
for $n_{i}$ number density of dark matter candidate species:  i=$\chi,\chi^{*},\eta_{2}$ 
and $n_{i}^{eq}$ their number density in thermal equilibrium with  background particle density $n=\sum_{i=1}^{N}n_{i}.$
Increasing the number of particle types increases the number of hidden sectors contributing to total relic density and the number of coupled Boltzmann equations we need to solve to find individual relic density of each sector. 
If masses of all particle types from hidden sectors are smaller than 1.5 times the mass of the lightest particle  the situation becomes simple since the number of Boltzmann equations reduces to one. In this case  i=$\chi,\chi^{*},\eta_{2},\eta_{3},\eta_{4}$.   It has been shown for a particular model \cite{Gondolo:1998ah} that co-annihilations can lower $\Omega h^{2}$ by factor of 6.
The condition (6) cannot be guaranteed when semi-annihilations are present \cite{D'Eramo:2010ep}.
This assumption that DM particles which contribute to co-annihilation processes are  in relative thermal equilibrium in the early universe  is also taken in the micrOMEGAs calculation.

\begin{figure}
 \moveleft0.11\linewidth\vbox{
        \includegraphics[width=1.16\linewidth]{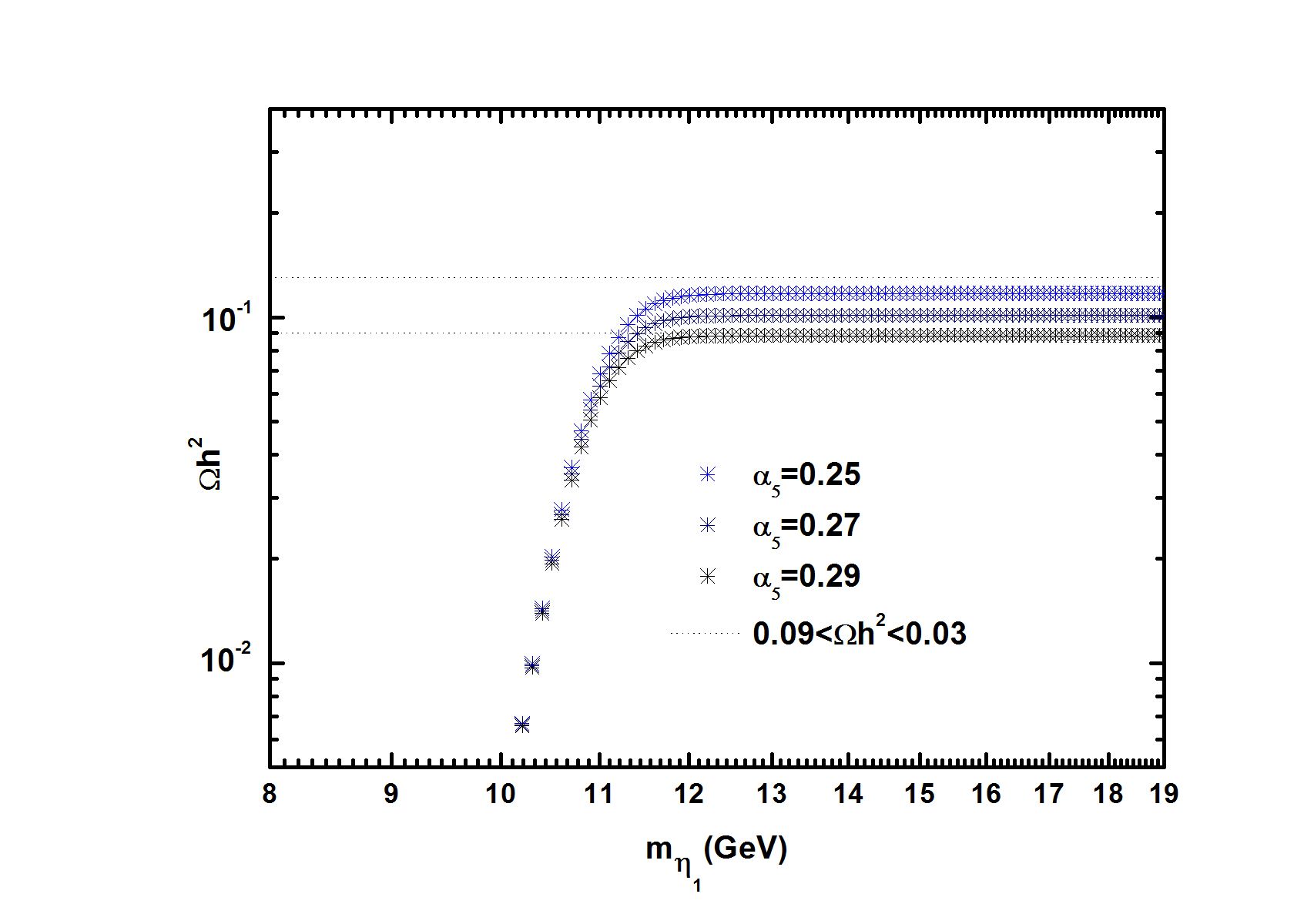}
    }
\caption{Dependence of the mass $m_{\eta_{1}}$ (second lightest particle) for $m_{\chi}=9\text{ GeV}$ on relic density.
Approaching  the mass of $\eta_{1}$  to the mass of the lightest particle results in contribution of  type (c) reactions. }
	\label{co}
\end{figure}

To investigate the influence of almost degenerate masses  in quaternion model we choose masses of DM particles very similar
as it is recommended for typical freeze-out temperature $T=m_{\chi}/20$,  while the recommended mass of the heavier particles, $\eta_{i}$ for i=2, 3, 4, is $m_{\eta_{i}}<1.5 m_{\chi}$ under which co-annihilations become included. 
 In the region of lower masses this effect is shown on the Figure \ref{co}.
Since $\Omega h^{2}$ depends on Higgs portal coupling we show the behavior of $\Omega h^{2}$  depending on the  mass of the second lightest DM particle approaching to mass of DM candidate for couplings $\alpha_{5}=0.25,0.27,$ and 0.29, determined from Figure \ref{alph5_mchi}.
The couplings which start to show effect are those with $\alpha_{4i}$ with i=2,3,4.
Dominant reactions  from the "dark" sector are those governed by Higgs channel.  Analysis shows that the effect of couplings $\alpha_{42}$, $\alpha_{43}$ and $\alpha_{44}$ is visible for the entire range of masses $m_{\chi}$ when coupling $\alpha_{5}$ is smaller than 0.001. 


\vspace{0.5cm}

\textbf{Direct detection}.
One of the ways to search for dark matter is measuring the nuclear recoil induced by DM scattering off nuclei. For the scalar interactions, when DM candidate couples to SM particles via Higgs boson we have spin-independent (SI) interactions. The best nuclei  for detecting scalar interactions are heavy ones because scalar interactions add coherently in the nucleus \cite{Belanger:2008sj}. 

\begin{figure}[ht!]

 \moveleft0.08\linewidth\vbox{
        \includegraphics[width=1.16\linewidth]{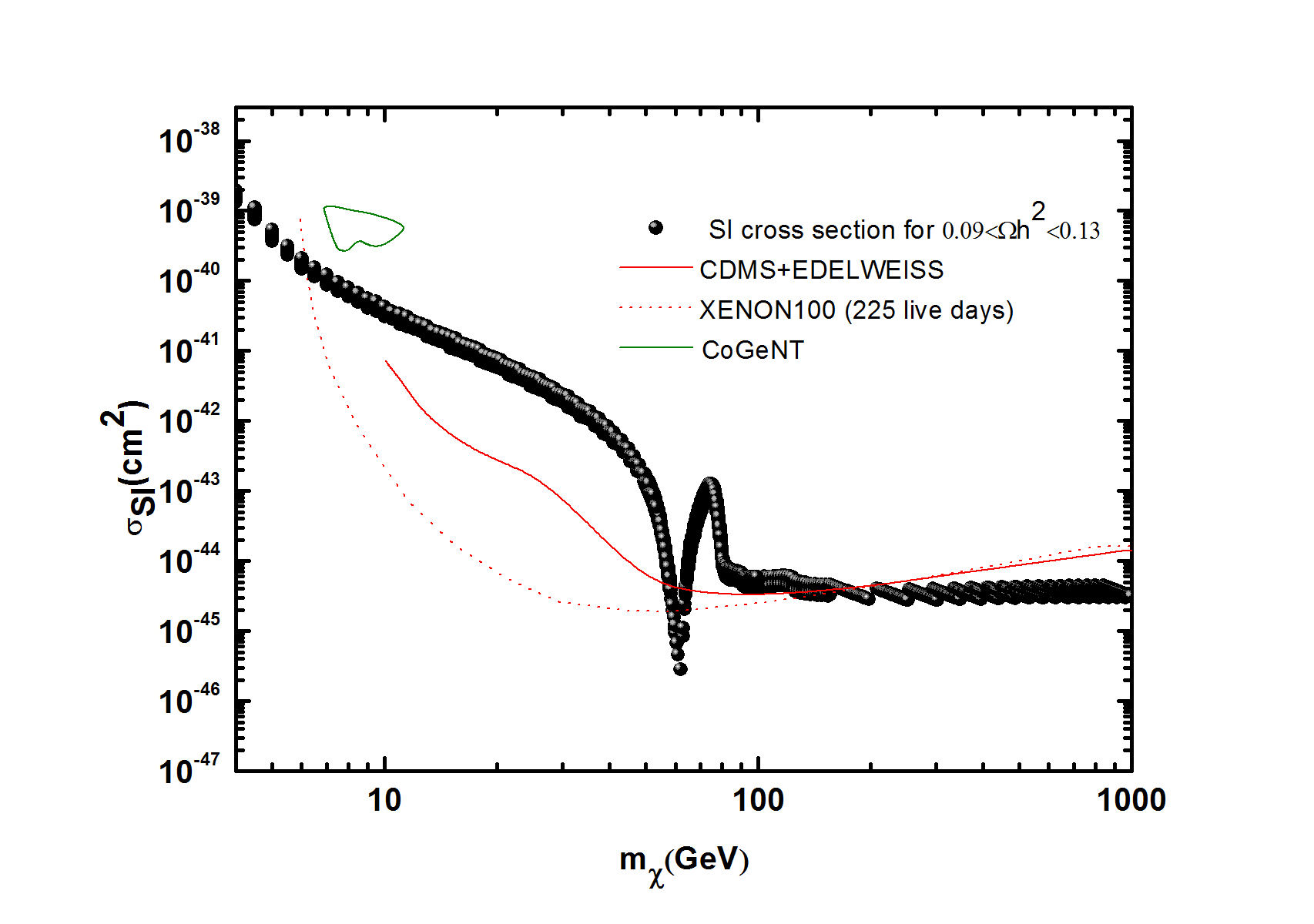}
    }
\caption{The figure shows spin-independent cross sections for reactions of (a) type, with $m_{\eta_{1}}=2100\text{ GeV}$ chosen to allow reactions (d) for every $m_{\chi}$.  The cross-sections (black) are obtained varying Higgs portal coupling $\alpha_{5}$ as shown on Figure \ref{alph5_mchi} to be consistent with  WMAP  measurements $0.09<\Omega h^{2}<0.13$. Full red line shows restriction from CDMS and EDELWEISS experimental results \protect{\cite{Ahmed:2011gh}}, the dotted one the results from XENON100 from 225 live days \protect{\cite{Aprile:2012nq}} and the green one is region investigated by CoGeNT. }
	\label{dir}
\end{figure}

There are many experiments searching for DM particles in this manner. DAMA \cite{Belli:2011kw}  has reported an excess in annual modulations effect   favoring light scalars as DM in agreement with a small excess of events by  CDMS, EDELWEISS  and CoGeNT   experiment  who have also reported a possible signal \cite{Ahmed:2011gh},\cite{Aalseth:2010vx}.  However, the null hypothesis of CoGeNT provides similar reduced chi square as  for dark matter particle of 9GeV with $\sigma_{SI}=6.7\cdot10^{-41}\text{ cm}^{2}$. Two events XENON100  observed in the region for WIMP search can as well be coming form  background fluctuations. The probability for this in the expected region  for WIMP search is 26.4\%.
They have provided the most stringent  limit for $m_{\chi}>8\text{ GeV}$ with a minimum of $\sigma=2.0\cdot10^{-45}\text{ cm}^{2}$ at $m_{\chi}=55\text{ GeV}$ \cite{Aprile:2012nq}.  Figure \ref{dir} shows the obtained spin-independent cross sections for annihilation of $\chi$ and $\chi^{*}$ particles for the case when we have annihilation reactions and decays.

 The parts of parameter space which are allowed under these conditions are heavy DM particles $m_{\chi}>165\text{ GeV}$, very light ones and (small) range of masses around Higgs resonance $\frac{m_{H}}{2}\approx60\text{ GeV}$. 
 CMS and EDELWEISS experiments restrict the higher masses as the XENON100 but allow masses around  the resonance.
Inclusion of additional reactions like semi-annihilations would provide a wider range of WMAP-restricted  $\alpha_{5}$ values and would possibly decrease the cross-section allowing a larger area of the parameter space. 
Light DM masses favored by CDMS, EDELWEIS, XENON100 and DAMA annual modulation can on the basis of direct detection measurements be accommodated in this model. 

\vspace{0.5cm}

\textbf{Indirect detection.} Primordial reactions which took place in the early Universe take place nowdays in regions assumed to consist of dense DM such as galactic center, "galactic halos or extra galactic region". Annihilation of DM in these regions results in a pair of SM particles which hadronize and decay into the stable particles and then propagate through Universe. 
When these particles are photons, neutrinos and anti-matter fluxes they can be detected by indirect detection experiments.  One of the sources  observed by Fermi-LAT are dwarf spheroidal galaxies (dSphs) of the Milky Way,  DM dominated systems, that makes them good targets for DM searches \cite{Ackermann:2011wa}.
Gamma ray signal from these measurements determines upper limits on DM annihilation cross section.
In determination of gamma ray flux (in number of photons per $cm^{2}s\textit{ sr}$)
\eq{\phi_{\gamma}(E,\phi)=\frac{\sigma v}{m_{\chi}^{2}}f_{\gamma}H(\phi),} model dependence enters in DM mass, annihilation cross section $\sigma v$ and annihilation rates in different annihilation channels. Here $f_{\gamma}(E)=\frac{d N_{\gamma}}{d E}$ is energy distribution of particle $\gamma$ produced in one reaction and $H(\phi)$ is integral of the dark matter density squared over the line of sight and dependent on the halo profile which is in our case Navarro- Frenk-White (NFW) profile \cite{Belanger:2010gh}.

\begin{figure}[ht!]
	
 \moveleft0.08\linewidth\vbox{
        \includegraphics[width=1.16\linewidth]{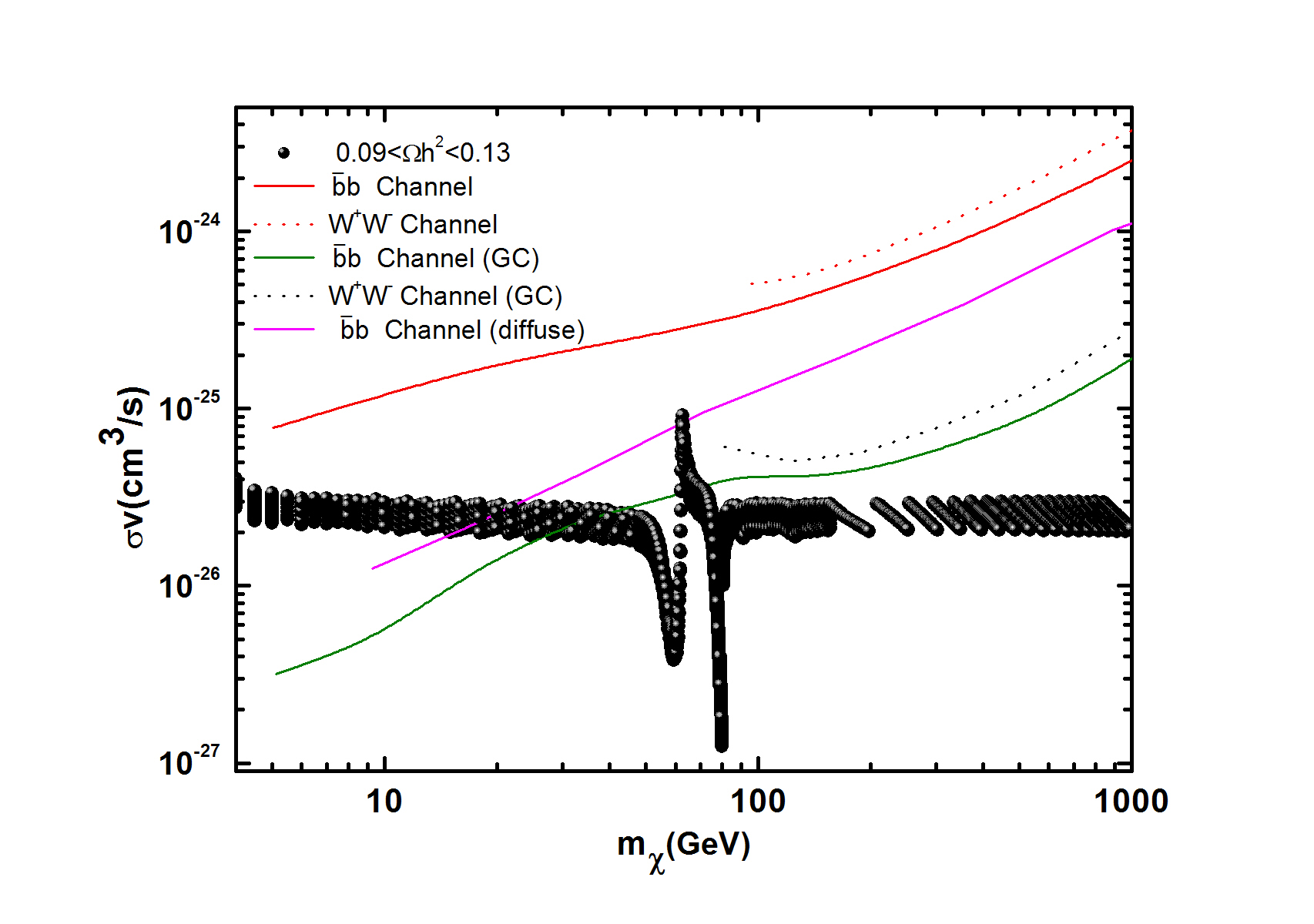}
    }
\caption{The figure shows annihilation cross section in  dependence on $m_{\chi}$ for the WMAP allowed values of $\alpha_{5}$ and values of $m_{\eta_{i}}=2100 \text{ GeV}$ for i=2,3,4. The red solid line shows upper limit for $\overline{b}b$  and dotted red line for $W^{+}W^{-}$ dominant annihilation channels measured by Fermi-LAT from observing dSphs and obtained using combined analysis \protect{\cite{Ackermann:2011wa}}. Green lines are upper bounds from measuring the gamma rays from Galactic Center (GC) for $\overline{b}b$ and $W^{+}W^{-}$ shown with solid and dotted line respectively \protect{\cite{Hooper:2012sr}}, while the pink line is obtained from diffuse gamma ray emission from Galactic Halo \protect{\cite{Ackermann:2012rg}}. }
	\label{ind}
\end{figure}

Figure \ref{ind} shows the $\left<\sigma v\right>$ as a function of $m_{\chi}$ for masses from 4 GeV to 1000 GeV. For small masses  $m_{\chi}<5\text{ GeV}$ dominant annihilation channel is into $c\overline{c}$  quarks  which for the masses above 5 GeV become $b\overline{b}$. This annihilation channel prevails until around 75 GeV after which $\chi$ particles dominantly annihilate into $W^{+}W^{-}$ bosons. 
We also show the upper limits measured by Fermi-LAT  observing dSphs with $b\overline{b}$  and $W^{+}W^{-}$ main annihilation channel with later one for heavier masses. For the mass range from 95 GeV to 1000 GeV they give upper bound on $\left<\sigma v\right>$ value  from the order of $\sim10^{-25}$ to the order of $\sim2\cdot 10^{-24}$ \cite{Ackermann:2011wa}. The most stringent constraints come from DM annihilation measurement from the region of Galactic Center (GC) by Fermi-LAT \cite{Hooper:2012sr}. They exclude the light masses $\leq20$ GeV of the model while the measurements of diffuse gamma ray emission from Galactic Halo (masking out Galactic Plane) exclude the masses lighter than around 15 GeV \cite{Ackermann:2012rg}.  We can notice that investigation of both GC and dSphs allows for heavier masses of the model.

Observations of Coma Cluster of Galaxies produced a little less stringent constraints on $\left<\sigma v\right>$. They are  of the order of $\sim10^{-21}$\text{, }$\sim10^{-20}$ \cite{Pfrommer:2012mm} in the lowest region of parameter space.

\section{Conclusion}

We  have shown that this very simple model based on non-Abelian finite group with one SM Higgs doublet
can with certain choice of parameters be very similar to $D_{3}$ model with one component DM. 
From the comparison with direct detection experiments we found that favorable  region of light DM is also allowed in $Q_{4}$ model. The large part of the model parameter space is very constrained while two other allowed areas are masses above order of 165 GeV  and masses around Higgs resonance.  The indirect detection experiments give little less stringent bounds allowing the masses above around 25 GeV. \newline
Quaternion model can, like any scalar DM model, be further investigated via invisibly decaying Higgs boson in  vector boson fusion (VBF) and ZH production channels at ATLAS experiment \cite{Gagnon:2009zz}. There is a possibility of searching for monochromatic gamma ray lines from annihilation of particles  via virtual SM Higgs into two gamma ray lines \cite{Adulpravitchai:2011ei}.
\newline
 We can conclude that allowed discrete non-Abelian group obtained from breaking of continuous SU(2) gauge symmetry can lead to new dark matter phenomena  and new observations. Findings  through that kind of new phenomena make steps toward unmasking DM nature.

\section{Acknowledgments}

We thank  Kre\v{s}imir Kumeri\v{c}ki,  Patrick O. Ludl  and Alexander Pukhov for useful comments and discussions.

\vspace{3cm}



\begin{thebibliography}{99}


\bibitem{Bergstrom:2012fi} 
  L.~Bergstrom,
  arXiv:1205.4882 [astro-ph.HE].


\bibitem{Blum:2007jz} 
  A.~Blum, C.~Hagedorn and M.~Lindner,
  Phys.\ Rev.\ D {\bf 77}, 076004 (2008)
  [arXiv:0709.3450 [hep-ph]].


\bibitem{Zhou:2012zj} 
  S.~Zhou,
  arXiv:1205.0761 [hep-ph].

\bibitem{Krauss:1988zc} 
  L.~M.~Krauss and F.~Wilczek,
  Phys.\ Rev.\ Lett.\  {\bf 62}, 1221 (1989).



\bibitem{Belyaev:2006zz} 
  A.~Belyaev, C.~-R.~Chen, K.~Tobe and C.~P.~Yuan,
  Conf.\ Proc.\ C {\bf 060726}, 1113 (2006).

\bibitem{Burgess:2000yq} 
  C.~P.~Burgess, M.~Pospelov and T.~ter Veldhuis,
  Nucl.\ Phys.\ B {\bf 619}, 709 (2001)
  [hep-ph/0011335].


\bibitem{Ma:2007gq} 
  E.~Ma,
  Phys.\ Lett.\ B {\bf 662}, 49 (2008)
  [arXiv:0708.3371 [hep-ph]].



\bibitem{Belanger:2012vp} 
  G.~Belanger, K.~Kannike, A.~Pukhov and M.~Raidal,
  JCAP {\bf 1204}, 010 (2012)
  [arXiv:1202.2962 [hep-ph]].

\bibitem{Adulpravitchai:2011ei} 
  A.~Adulpravitchai, B.~Batell and J.~Pradler,
  Phys.\ Lett.\ B {\bf 700}, 207 (2011)
  [arXiv:1103.3053 [hep-ph]].


\bibitem{Frigerio:2004jg} 
  M.~Frigerio, S.~Kaneko, E.~Ma and M.~Tanimoto,
  Phys.\ Rev.\ D {\bf 71}, 011901 (2005)
  [hep-ph/0409187].

\bibitem{Frigerio:2005pz} 
  M.~Frigerio,
  hep-ph/0505144.

\bibitem{Frigerio:2007nn} 
  M.~Frigerio and E.~Ma,
  Phys.\ Rev.\ D {\bf 76}, 096007 (2007)
  [arXiv:0708.0166 [hep-ph]]. 	 	

\bibitem{Adulpravitchai:2009kd} 
  A.~Adulpravitchai, A.~Blum and M.~Lindner,
  JHEP {\bf 0909}, 018 (2009)
  [arXiv:0907.2332 [hep-ph]].

\bibitem{Ishimori:2010au} 
  H.~Ishimori, T.~Kobayashi, H.~Ohki, Y.~Shimizu, H.~Okada and M.~Tanimoto,
  Prog.\ Theor.\ Phys.\ Suppl.\  {\bf 183}, 1 (2010)
  [arXiv:1003.3552 [hep-th]].

\bibitem{Grimus:2011fk} 
  W.~Grimus and P.~O.~Ludl,
  J.\ Phys.\ A A {\bf 45}, 233001 (2012)
  [arXiv:1110.6376 [hep-ph]].

\bibitem{Belanger:2006is} 
  G.~Belanger, F.~Boudjema, A.~Pukhov and A.~Semenov,
  Comput.\ Phys.\ Commun.\  {\bf 176}, 367 (2007)
  [hep-ph/0607059].

\bibitem{Belanger:2010gh} 
  G.~Belanger, F.~Boudjema, P.~Brun, A.~Pukhov, S.~Rosier-Lees, P.~Salati and A.~Semenov,
  Comput.\ Phys.\ Commun.\  {\bf 182}, 842 (2011)
  [arXiv:1004.1092 [hep-ph]].

\bibitem{Belanger:2008sj} 
  G.~Belanger, F.~Boudjema, A.~Pukhov and A.~Semenov,
  Comput.\ Phys.\ Commun.\  {\bf 180}, 747 (2009)
  [arXiv:0803.2360 [hep-ph]].

\bibitem{Angle:2009xb} 
  J.~Angle {\it et al.}  [XENON10 Collaboration],
  Phys.\ Rev.\ D {\bf 80}, 115005 (2009)
  [arXiv:0910.3698 [astro-ph.CO]].

\bibitem{Aprile:2010um} 
  E.~Aprile {\it et al.}  [XENON100 Collaboration],
  Phys.\ Rev.\ Lett.\  {\bf 105}, 131302 (2010)
  [arXiv:1005.0380 [astro-ph.CO]].

\bibitem{D'Eramo:2010ep} 
  F.~D'Eramo and J.~Thaler,
  JHEP {\bf 1006}, 109 (2010)
  [arXiv:1003.5912 [hep-ph]].

\bibitem{Boucenna:2011tj} 
  M.~S.~Boucenna, M.~Hirsch, S.~Morisi, E.~Peinado, M.~Taoso and J.~W.~F.~Valle,
  JHEP {\bf 1105}, 037 (2011)
  [arXiv:1101.2874 [hep-ph]].

\bibitem{Gondolo:1998ah} 
  P.~Gondolo and J.~Edsjo,
  Phys.\ Atom.\ Nucl.\  {\bf 61}, 1081 (1998)
  [Yad.\ Fiz.\  {\bf 61}, 1181 (1998)].

\bibitem{Belli:2011kw} 
  P.~Belli, R.~Bernabei, A.~Bottino, F.~Cappella, R.~Cerulli, N.~Fornengo and S.~Scopel,
  Phys.\ Rev.\ D {\bf 84}, 055014 (2011)
  [arXiv:1106.4667 [hep-ph]].

\bibitem{Ahmed:2011gh} 
  Z.~Ahmed {\it et al.}  [CDMS and EDELWEISS Collaborations],
  Phys.\ Rev.\ D {\bf 84}, 011102 (2011)
  [arXiv:1105.3377 [astro-ph.CO]].

\bibitem{Aalseth:2010vx} 
  C.~E.~Aalseth {\it et al.}  [CoGeNT Collaboration],
  Phys.\ Rev.\ Lett.\  {\bf 106}, 131301 (2011)
  [arXiv:1002.4703 [astro-ph.CO]].



\bibitem{Aprile:2012nq} 
  E.~Aprile {\it et al.}  [XENON100 Collaboration],
  arXiv:1207.5988 [astro-ph.CO].


\bibitem{Ackermann:2011wa} 
  M.~Ackermann {\it et al.}  [Fermi-LAT Collaboration],
  Phys.\ Rev.\ Lett.\  {\bf 107}, 241302 (2011)
  [arXiv:1108.3546 [astro-ph.HE]].

\bibitem{Hooper:2012sr} 
  D.~Hooper, C.~Kelso and F.~S.~Queiroz,
  arXiv:1209.3015 [astro-ph.HE].

\bibitem{Ackermann:2012rg} 
  T.~F.~-: M.~Ackermann {\it et al.}  [LAT Collaboration],
  arXiv:1205.6474 [astro-ph.CO].

\bibitem{Pfrommer:2012mm} 
  w.~C.~Pfrommer {\it et al.}  [The Veritas Collaboration],
  Astrophys.\ J.\  {\bf 757}, 123 (2012)
  [arXiv:1208.0676 [astro-ph.HE]].

\bibitem{Abdo:2010dk} 
  A.~A.~Abdo {\it et al.}  [Fermi-LAT Collaboration],
  JCAP {\bf 1004}, 014 (2010)
  [arXiv:1002.4415 [astro-ph.CO]].


\bibitem{Gagnon:2009zz} 
  P.~Gagnon {\it et al.}  [ATLAS Collaboration],
  ATL-PHYS-PUB-2009-061.


\end{thebibliography}

\end{document}